\documentclass[runningheads]{llncs}
\usepackage[T1]{fontenc}
\usepackage{listings}
\usepackage{makeidx}
\usepackage{algorithm, algpseudocode}
\usepackage{graphicx}
\usepackage{float}
\usepackage{cite}
\usepackage{amsmath}
\usepackage{rotating,todonotes, xspace}
\usepackage{multirow}
\usepackage{hyperref}
\usepackage{xspace}
\usepackage{paralist}
\usepackage{booktabs}
\usepackage{balance}
\usepackage{cleveref}
\usepackage{subcaption}
\usepackage{placeins}

\newcommand{\pp}[1]{\vspace{6pt}\noindent\textbf{\emph{#1.}}\xspace}

\begin{document}

\title{BanditWare: A Contextual Bandit-based Framework for Hardware Prediction}


\author{Tain\~a Coleman\inst{1}\orcidID{0002-0982-1919} \and
Hena Ahmed\inst{1}\orcidID{0003-4532-0334} \and
Ravi Shende\inst{1} \and Ismael Perez\inst{1} \and \.{I}lkay Altinta\c{s}\inst{1}\orcidID{0002-2196-0305}}

\authorrunning{T. Coleman et al.}

\institute{University of California San Diego, La Jolla, CA 92093, USA \\
\email{\{t1coleman,h7ahmed,rshende,i3perez,ialtintas\}@ucsd.edu}}


\maketitle

\begin{abstract}
Distributed computing systems are essential for meeting the demands of modern applications, yet transitioning from single-system to distributed environments presents significant challenges. Misallocating resources in shared systems can lead to resource contention, system instability, degraded performance, priority inversion, inefficient utilization, increased latency, and environmental impact.

We present BanditWare, an online recommendation system that dynamically selects the most suitable hardware for applications using a contextual multi-armed bandit algorithm. BanditWare balances exploration and exploitation, gradually refining its hardware recommendations based on observed application performance while continuing to explore potentially better options. Unlike traditional statistical and machine learning approaches that rely heavily on large historical datasets, BanditWare operates online, learning and adapting in real-time as new workloads arrive.

We evaluated BanditWare on three workflow applications: Cycles (an agricultural science scientific workflow) BurnPro3D (a web-based platform for fire science) and a matrix multiplication application. Designed for seamless integration with the National Data Platform (NDP), BanditWare enables users of all experience levels to optimize resource allocation efficiently.
\end{abstract}



\section{Introduction}
\label{sec-introduction}

Advancements in distributed and high-performance computing (HPC) have made resource allocation a critical research topic. However, only a small portion of this research has evolved into practical tools, leaving a gap between theory and application~\cite{shealy2021intelligent}. This gap is particularly significant as modern platforms grow increasingly complex and heterogeneous, posing challenges for domain scientists who may lack the computational expertise to make effective resource allocation decisions. Misallocating resources can severely impact application performance and the overall distributed system~\cite{shukur2020cloud}.

The increasing diversity of computing architectures has intensified these challenges, requiring new strategies for scalability and reliability. The US Department of Energy’s 2018 report on Extreme Heterogeneity~\cite{osti_1473756} emphasizes the need for advanced performance modeling and prediction to enable better resource allocation across diverse workflows. Additionally, the rapid development of AI/ML and AI-for-science workflows, which are data- and compute-intensive, has highlighted the need for efficient resource allocation\cite{emani2022ai}, especially for urgent computing workflows like weather forecasting and emergency response. Misallocating in these contexts can lead to resource contention, system instability, increased latency, and significant financial costs.




This paper addresses the problem of identifying the best-suited hardware setting for workflows with respect to execution time. We present \textbf{BanditWare}, an \textit{online learning}, self-updating recommendation system that achieves measurably sufficient recommendation accuracy using minimal iterations and data samples. We evaluate the utility of BanditWare for the National Data Platform (NDP)~\cite{NDP}, which is a federated and extensible data ecosystem that promotes collaboration, innovation, and equitable use of data over existing cyberinfrastructure (CI) capabilities. We use the following NDP application workflows as use cases for BanditWare in this study:


\pp{Cycles~\cite{da2019empowering}} A Scientific workflow that describes an Agroecosystem model for simulations of crop production and natural element cycles in the soil-plant-atmosphere space. 

\pp{BurnPro3D (BP3D)~\cite{bp3d}} A web-based platform that combines next-generation fire science, data, and artificial intelligence (AI) to optimize prescribed burns on the scale needed to reduce wildfire risk and meet land objectives by using a data- and simulation-driven approach.

\pp{Matrix Multiplication} A fully parallelized, tiled matrix squaring algorithm that takes advantage of the full number of CPU cores given to it, allowing it to perform differently on varied hardware configurations. \\

We selected NDP as a testbed because of its heterogeneous, geo-distributed Kubernetes cluster spanning the United States~\cite{nrp2008}. BanditWare's rapid recommendation capability benefits platforms like NDP by empowering domain scientists—regardless of their computational expertise—to select the appropriate hardware for their experiments. This framework is an important step towards enhancing performance, reducing resource waste, and ensuring that workflows execute in optimal environments.


The remainder of this paper is organized as follows: Section~\ref{sec-related-work} summarizes related work, Section~\ref{sec-approach} describes the approach and methodology of our study, Section~\ref{sec-experiments} contains the use-case results and evaluation of BanditWare, and Section~\ref{sec-conclusion} presents conclusions and future work.

\section{Related Work}
\label{sec-related-work}

Over the past decades, distributed systems have grown increasingly complex and heterogeneous to meet the rising demand for fast and reliable large-scale computational workflows~\cite{jacob2009montage} and AI/ML applications~\cite{imagenet, achiam2023gpt}. The study of workloads focused on improving scheduling and resource management during such expensive executions, often by creating models for system analysis or simulation. Lublin et al.~\cite{LUBLIN20031105} presented an intrinsic model for parallel jobs with respect to sizes, runtime, and arrival times. Feitelson~\cite{workload_modeling} extended predictive modeling workflows to include user behavior. Lee et al.~\cite{lee2023case} proposed a framework to optimize resource costs for data management tasks in ML-for-science workflow applications on heterogeneous distributed systems. However, these studies focused exclusively on offline approaches for resource allocation, and as Emeras et al.~\cite{emeras2014analysis} showed, allocation estimates often deviated from actual usage. To achieve more efficient scheduling and resource management, both resource allocation and usage needed to be understood.

Tools like the semi-automated, platform-agnostic Tesseract~\cite{shealy2021intelligent} addressed these issues by predicting resource usage. Evalix~\cite{evalix} similarly improved the matching of scheduled jobs to available resources while reducing energy costs. Other tools, such as the Hardware Locality (hwloc) software~\cite{goglin2014managing}, enabled users to map the topology of available hardware resources, aiding task assignment in distributed, heterogeneous computing platforms.

High-performance computing (HPC) researchers have increasingly adopted artificial intelligence and machine learning (AI/ML) methods to enhance inferential capabilities. Nassereldine et al.~\cite{nassereldine2023tradeoff} described a performance prediction tool that trained supervised learning models to evaluate trade-offs between computing costs and the performance of distributed HPC workflows. Mastanuga and Fortes~\cite{matsunaga2010use} similarly framed application resource consumption prediction as a supervised machine learning problem, presenting an experimental use case that considered both system performance and application-specific attributes in heterogeneous hardware environments. Recent work built on this foundation, integrating AI/ML predictive modeling into resource optimization and provisioning methods. Emani et al.~\cite{emani2022ai} evaluated the impact of AI hardware accelerators on meeting resource consumption needs for scientific ML/DL workloads. AI/ML predictive tools also advanced in data-driven applications such as recommendation systems~\cite{portugal2018use, afsar2022reinforcement}. Bobadilla et al.~\cite{bobadilla2013} introduced HwSimilarity, a similarity metric designed to improve the speed of k-nearest neighbor algorithms used in recommendation systems. This metric reduced computational time while maintaining high-quality recommendations, achieving a significant speed increase by simplifying computations into a binary format suitable for Boolean functions. Bhareti et al.~\cite{bhareti2020} provided a comprehensive review of recommendation system evolution, highlighting hybrid models that integrated machine learning, neural networks, and deep learning to deliver accurate and reliable results.

Ahmed et al.~\cite{ahmed2024integratedperformanceframeworkscience} proposed a performance prediction framework integrating data flows, ML/AI workflows, and application services. Their use case demonstrated predictive ML/AI modeling to inform resource provisioning for intensive application workloads, including simulations on the BurnPro3D platform.

Unlike these works, our framework emphasizes lightweight, online decision-making. While approaches like Ahmed et al.~\cite{ahmed2024integratedperformanceframeworkscience} rely on extensive offline analyses, our work focuses on adapting to dynamic environments through real-time responsiveness. This enables practical deployment in scenarios requiring minimal historical data and iterative refinement.

\section{Approach}
\label{sec-approach}

This section outlines the problem objective and the methodology behind the design of BanditWare. We describe the components involved in achieving the best-fit hardware recommendation. Section~\ref{subsec-overview} presents an overview of the framework and explains how these components work together and Section~\ref{subsec-alg} describes the contextual bandit algorithm used to determine the hardware recommendation.

\pp{Problem Statement}
\label{subsec-problem}

Optimizing application performance through appropriate hardware selection is a critical problem in today's rapidly evolving technological landscape. Applications often vary in their computational requirements and execution profiles, making it essential to match them with the most suitable hardware to ensure efficiency, cost-effectiveness, and improved performance. The key challenges of this problem are:

\begin{itemize}
    \item \textbf{Diverse Hardware Configurations}: The wide range of available hardware options, including CPUs, GPUs, and memory architectures, complicates the decision-making process.
    \item \textbf{Variable Application Workloads}: Applications vary significantly in their computational and I/O requirements, making it difficult to generalize hardware recommendations.
    \item {\textbf{Variable Performance Improvements}: Certain applications, such as parallelized or memory-intensive applications, show significantly greater performance gains when using specific hardware configurations compared to others.}
    \item \textbf{Data Analysis}: Efficiently processing and analyzing large volumes of execution data to derive meaningful insights and recommendations.
    \item {\textbf{Data Availability}: Scientific workflows can be large (on the scale of thousands of tasks), and collecting real trace data from these applications can be costly. This lack of historical data makes it challenging for complex traditional ML/AI prediction models to perform well in such applications.}
\end{itemize}

Given these challenges, we propose BanditWare, a comprehensive framework that leverages application execution data to recommend the best hardware configurations based on system availability. The proposed framework uses multi-armed bandit and data analysis techniques to predict optimal hardware setups, aiming to facilitate better resource utilization and improved performance by minimizing the application's overall runtime.

\subsection{Overview of BanditWare}
\label{subsec-overview}

\begin{figure}[ht!]
    \centering
    \includegraphics[trim={2cm 1cm 14cm 4.5cm},clip,width=\linewidth]{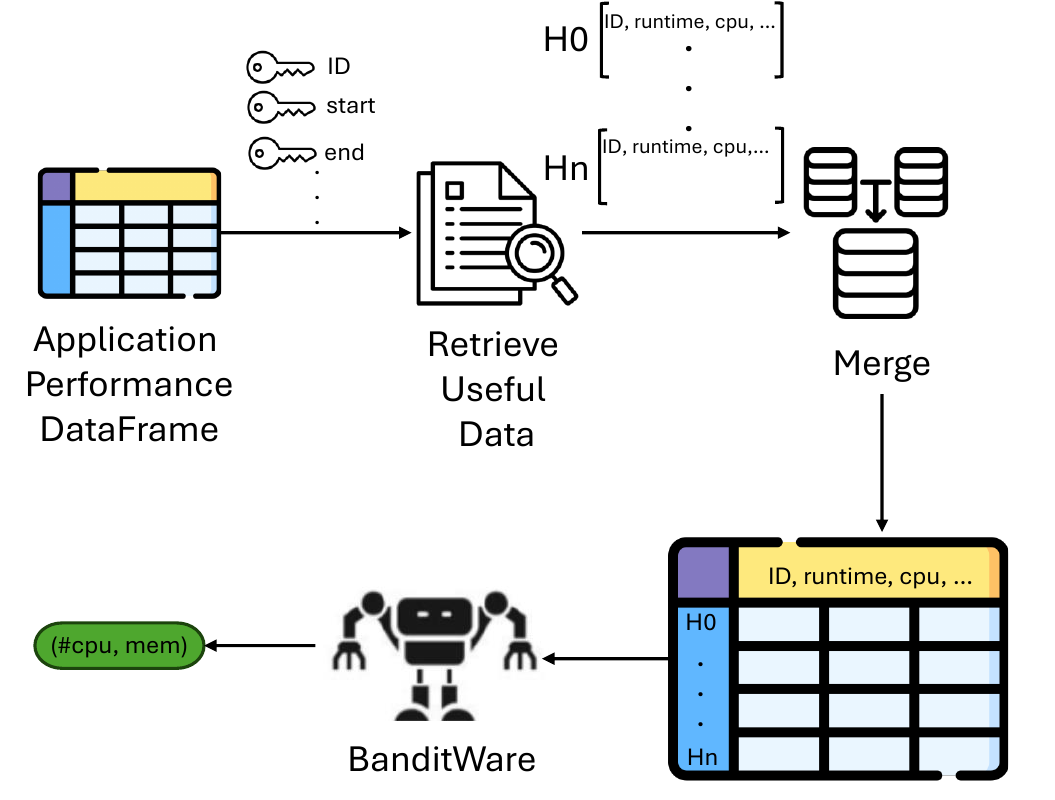}
    \caption{Overview of our framework.}
    \label{fig:framework}
\end{figure}

Figure~\ref{fig:framework} presents the overview of our framework. We start with a small dataset of application runs collected previously~\cite{ahmed2024integratedperformanceframeworkscience} which contains the application's input/output information, ID, runtime, start time, etc. The data is offered as input to BanditWare as a \texttt{Python pandas dataframe} where we parse the input features and runtimes per hardware to perform our experiments. 

\subsection{Algorithm}
\label{subsec-alg}

\begin{figure}[h]
    \centering
    \includegraphics[trim={.5cm 1.5cm .3cm 0cm},clip,width=0.7\linewidth]{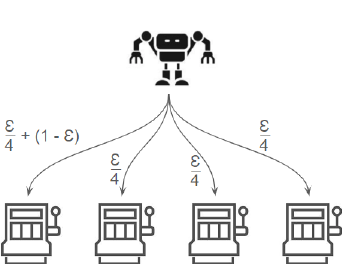}
    \caption{A multi-armed bandit problem using the $\epsilon$-greedy algorithm.}
    \label{fig:mab}
\end{figure}

The algorithm behind the BanditWare recommendation framework is based on a class of algorithms known as \emph{Multi-Armed Bandit Algorithms}. Although simple, these algorithms are highly effective when the problem involves making decisions over time and under uncertainty~\cite{slivkins2019introduction}. The term "multi-armed bandit" originates from a problem where a person plays several slot machines, also known as one-armed bandits, in a casino. The gambler can pull only one arm per round. Each machine has a different potential return or probability of winning, which the gambler does not know. The goal of the gambler is to maximize their reward, that is, to win as much money as possible~\cite{mab_jacko}. In each round, the player faces two choices: (i) continue playing the arm they previously chose, or (ii) select a different arm, with the possibility that it could either be better (more likely to win) or worse. This decision is known as the exploration/exploitation trade-off in learning algorithms. A simple but effective multi-armed bandit algorithm is the $\epsilon$-greedy strategy~\cite{context_bandit_1}.
Figure~\ref{fig:mab} illustrates a problem instance using $\epsilon$-greedy algorithm.

Various multi-armed bandit algorithms, each offering distinct advantages and disadvantages for different types of problems, have been proposed in the literature~\cite{zhou2015survey_bandit}. These algorithms are based on different mathematical models, such as Bayesian bandits, Lipschitz bandits, and linear cost bandits. The choice of the most appropriate algorithm depends on the problem's specific characteristics. 
In this work, we employ a \emph{Decaying Contextual $\epsilon$-Greedy Strategy}, where the exploration probability $\epsilon$ decreases over time, balancing exploration and exploitation based on available context (features of the workflows). The algorithm also incorporates parameters \texttt{tolerance\_ratio} and \texttt{tolerance\_seconds}, which allow trade-offs between runtime optimization and minimizing hardware resource usage. These parameters can be set to zero if runtime optimization is prioritized.

The algorithm assumes that workflows are represented by an $m$-dimensional feature vector $\mathbf{x} \in \mathbf{R}^m$ and that the runtime of a workflow on a given hardware $H_i$ follows a linear model: 
\[
R(H_i, \mathbf{x}) = \mathbf{w}_i^T \mathbf{x} + b_i
\]
where $\mathbf{w}_i$ and $b_i$ are unknown coefficients specific to hardware $H_i$.

\begin{algorithm}[H]
    \caption{Decaying Contextual $\epsilon$-Greedy Strategy with Tolerant Selection}
    \label{alg:mab}
    \begin{algorithmic}[1]
        \Require Hardware set $\mathcal{H} = \{H_0, H_1, ..., H_n\}$, workflows $\mathcal{W}$
        \Require Decay factor $\alpha$, initial exploration rate $\epsilon_0$
        \Require Tolerance parameters: tolerance ratio $t_r$, tolerance seconds $t_s$
        \State Initialize data storage: $\mathcal{D}_i \gets \emptyset, \forall i$
        \State Initialize coefficients: $\mathbf{w}_i \gets \mathbf{0}, \forall i$, bias $b_i \gets 0, \forall i$
        \State Set exploration rate: $\epsilon \gets \epsilon_0$
        \For{each incoming workflow $w_j \in \mathcal{W}$ with feature vector $\mathbf{x}_j$}
            \State Compute estimated runtime for each hardware:
            \[
            \hat{R}(H_i, \mathbf{x}_j) = \mathbf{w}_i^T \mathbf{x}_j + b_i, \quad \forall H_i \in \mathcal{H}
            \]
            \State With probability $\epsilon$, select hardware randomly from $\mathcal{H}$ (exploration).
            \State With probability $1 - \epsilon$, select hardware using a tolerant selection strategy:
            \begin{itemize}
                \item Identify hardware with the minimum estimated runtime: \( H_{\text{fastest}} \)
                \item Compute tolerance threshold: 
                \[
                R_{\text{limit}} = (1 + t_r) \cdot \hat{R}(H_{\text{fastest}}, \mathbf{x}_j) + t_s
                \]
                \item Among hardware satisfying \( \hat{R}(H_i, \mathbf{x}_j) \leq R_{\text{limit}} \), choose the one with the most resource efficiency.
            \end{itemize}
            \State Schedule workflow $w_j$ on the selected hardware $H_k$.
            \State Observe actual runtime $R_{\text{actual}}(H_k, w_j)$.
            \State Store observed data for $H_k$: 
            \[
            \mathcal{D}_k \gets \mathcal{D}_k \cup \{ (\mathbf{x}_j, R_{\text{actual}}) \}
            \]
            \State Perform Least Squares regression using stored data $\mathcal{D}_k$:
            \[
            \mathbf{w}_k, b_k = \arg\min_{\mathbf{w}, b} \sum_{(\mathbf{x}, R) \in \mathcal{D}_k} (R - (\mathbf{w}^T \mathbf{x} + b))^2
            \]
            \State Decay exploration rate: $\epsilon \gets \alpha \cdot \epsilon$
        \EndFor
    \end{algorithmic}
\end{algorithm}

This algorithm ensures that as more workflows are scheduled and executed, the model continuously refines its predictions of hardware performance, leading to better recommendations over time. The use of a decaying $\epsilon$-greedy strategy balances exploration and exploitation dynamically, allowing the system to efficiently adapt to new data.

\section{Experiments}
\label{sec-experiments}





For our experiments, we evaluate Banditware for three different workflow applications:

\begin{enumerate}
    \item \pp{Cycles} is an Agroecosystem model for simulations of crop production and water, carbon (C), and nitrogen (N) cycles in the soil-plant-atmosphere continuum. The model simulates agronomic practices and perturbations of the biochemical processes caused by them.  

    \item \pp{BurnPro3D} BurnPro3D (BP3D) is a physics-based simulation platform for prescribed fires. It uses GeoJSON files, known as burn units, to represent the geographic area of a prescribed burn and 3D vegetation data. For our experiments, we chose six burn units from previous simulations, ensuring that they were relevant to real-world data. The selection included burn units of varying sizes and regions to provide a more comprehensive dataset. 
    We conducted simulations for each burn unit, varying the weather input conditions, and repeated the process across all hardware configurations to create a well-rounded dataset.


    \item \pp{Matrix Multiplication} We chose a simple matrix squaring algorithm to test BanditWare's capabilities on an application that is more sensitive to the hardware it runs on. The input matrices consist of randomly generated integers with several parameters, although matrix generation is not included in the runtime measurement. The matrix size is the most highly correlated input parameter with runtime. In addition to size, we also vary sparsity (the ratio of zeros in the matrix) and the minimum and maximum values used for random number generation in the matrix.

\end{enumerate}

For all of our experiments, we run Algorithm~\ref{alg:mab} with a decay factor $\alpha = 0.99$ and initial exploration probability $\epsilon_0 = 1$. For the experiments on sections~\ref{sec-experiments-bp3d} and~\ref{sec-experiments-mm} we ran our applications on different hardware configurations described by $Hn=(\#\text{cpus}, \text{memory})$. We selected three hardware settings from those available on the open-source NDP: $H0 = (2, 16)$, $H1 = (3, 24)$, and $H2 = (4, 16)$.

\subsection{Experiment 1: Cycles on Synthetic Hardware}\label{sec-experiments-cycles}

\begin{figure}[h!]
    \centering
    \includegraphics[trim={.2cm 0.3cm 0.4cm 0cm},clip,width=\linewidth]{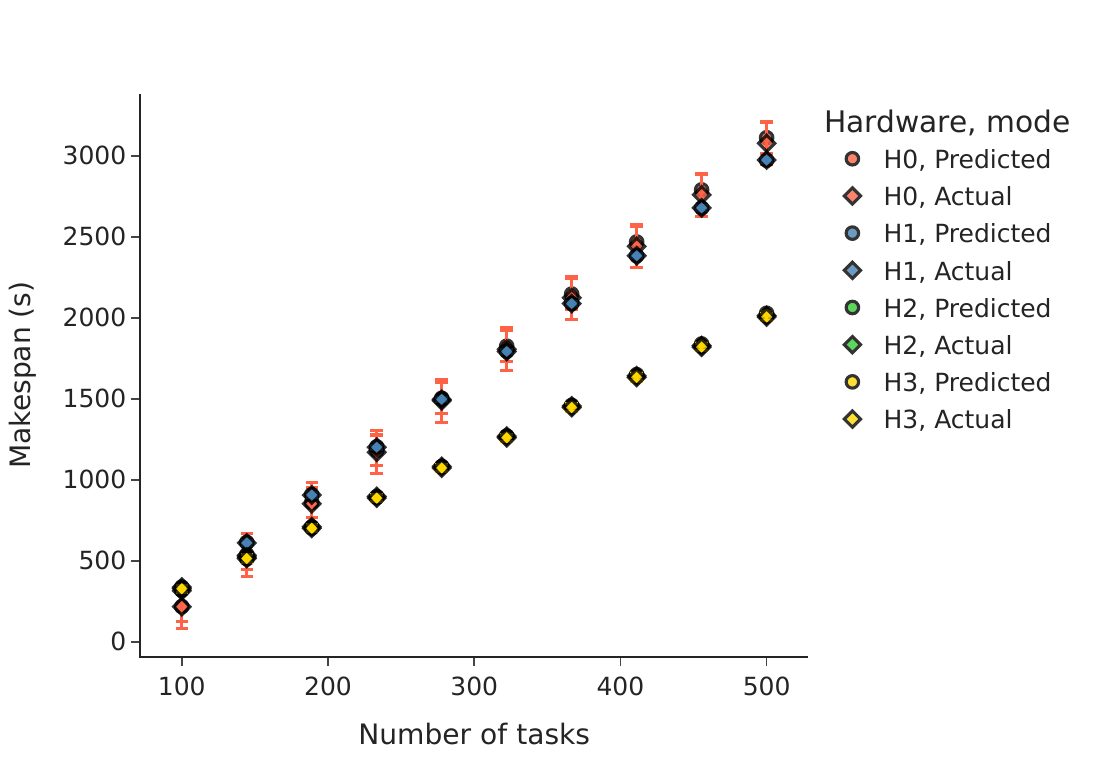}
    \caption{Each color represents a hardware configuration. The plot contains the actual fitting of the data represented by the diamond-shaped markers and our model fitting as the circle marker.}
    \label{fig:motivation}
\end{figure}

We analyzed data from 80 runs of the Cycles~\cite{9006107} workflow application of two different sizes: 100 tasks and 500 tasks. Figure~\ref{fig:motivation} presents the linear fitting for four different synthetic hardware settings, where the number of tasks (\texttt{num\_tasks}) is the only input feature.


\begin{figure}[h!]
    \centering
    \begin{subfigure}[b]{.48\textwidth}
        \centering
        \includegraphics[trim={.2cm .3cm 2cm 0.4cm},clip,width=\linewidth]{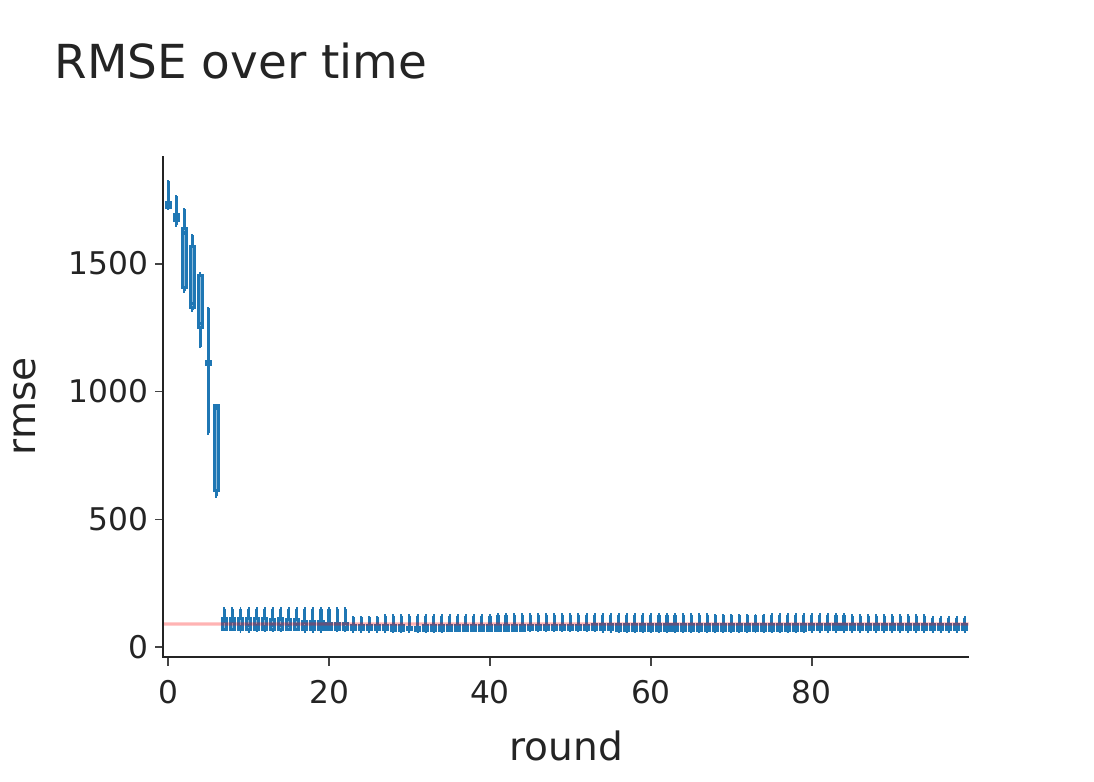}
        \caption{Root Mean Squared Error (RMSE) over time. The red line shows performance on the full dataset, while the blue bars represent variation across 10 simulations at each round.}
        \label{fig:rmse}
    \end{subfigure}%
    \hfill
    \vspace{2mm}
    \begin{subfigure}[b]{.48\textwidth}
        \centering
        \includegraphics[trim={.2cm .3cm 2cm .4cm},clip,width=\linewidth]{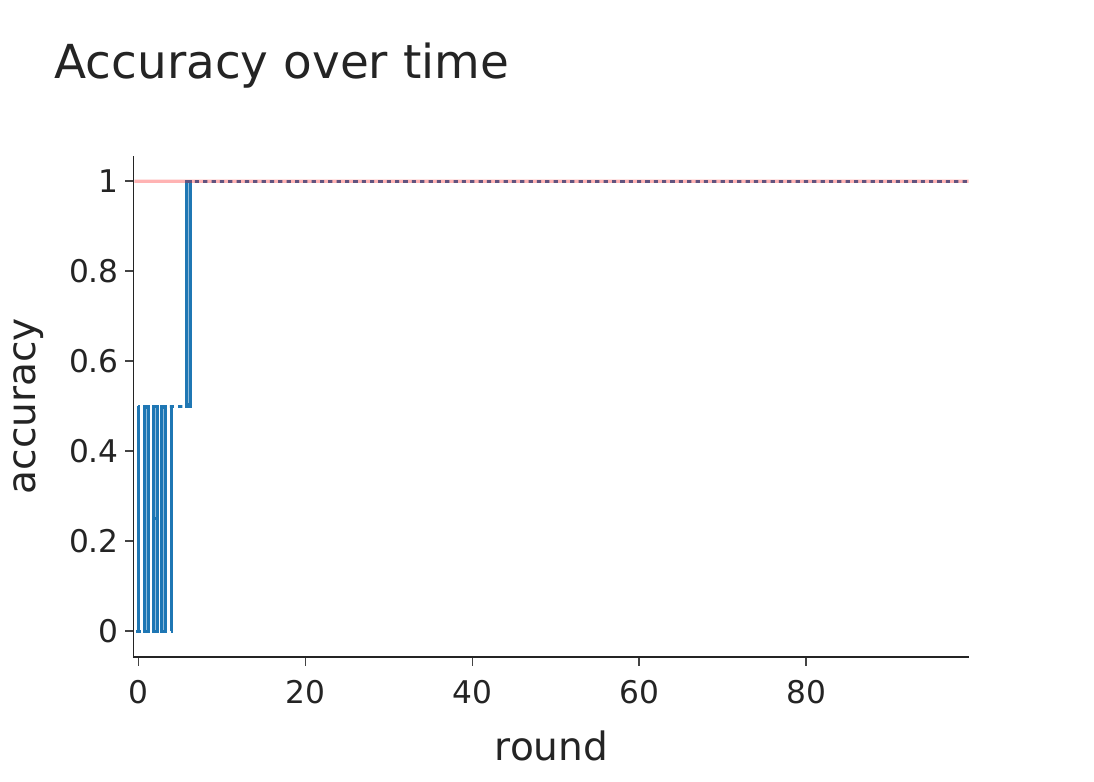}
        \caption{Accuracy over time. The red line shows results on the full dataset, while the blue bars represent variation across 10 simulations at each round.}
        \label{fig:accuracy}
    \end{subfigure}

    \caption{RMSE and accuracy of our model over time, comparing full dataset performance (red line) with variations across 10 simulations per round (blue bars).}
    \label{fig:rmse_accuracy}
\end{figure}

Figure~\ref{fig:rmse} shows the RMSE (Root Mean Square Error) over the 100 rounds of our model. We observe that Algorithm~\ref{alg:mab} achieves the same error rate as using 1316 data points (indicated by the horizontal red line), but with only 20 samples (approximately 98.5\% fewer data points). 
We also evaluated the accuracy of our model in predicting the best-fitting hardware for the entire dataset, using a tolerance of 20 seconds (see Figure~\ref{fig:accuracy}). 


These results suggest that when the runtime can be predicted as a linear combination of input variables and the hardware configurations present a meaningful trade-off, Algorithm~\ref{alg:mab} can learn an effective model for the hardware in just a few rounds.


\subsection{Experiment 2: BP3D on NDP Hardware}\label{sec-experiments-bp3d}

The dataset for this experiment comes from BP3D simulation runs on different Kubernetes configurations (which we refer to as hardware settings throughout this paper). Table~\ref{tab:bp3d-inputs} presents the input features for  BP3D workflows that we consider in this experiment.

\begin{table}[h!]
  \centering
  \caption{BurnPro3D Inputs \& Outputs}
  \begin{tabular}{|cl|}
    \hline
    \textbf{Feature Name} & \textbf{Description} \\
    \hline
    \texttt{surface\_moisture} & surface fuel moisture \\
    \texttt{canopy\_moisture} & canopy fuel moisture \\
    \texttt{wind\_direction} & direction of surface winds \\
    \texttt{wind\_speed} & speed of surface winds \\
    \texttt{sim\_time} & maximum simulation steps allowed\\
    \texttt{run\_max\_mem\_rss\_bytes} & maximum RSS bytes allowed per run \\
    \texttt{area} & calculated regional surface area\\
    \hline
  \end{tabular}
  \label{tab:bp3d-inputs}
\end{table}

We know from prior work~\cite{ahmed2024integratedperformanceframeworkscience} that the BP3D workflow runtime can be accurately modeled as a linear combination of the input features. We first present the results obtained by applying a linear regression recommender over small subsets of the full historical data. The goal is to compare the effectiveness of our approach with common recommendation algorithms.





\begin{figure}[h!]
    \centering
    \begin{subfigure}[b]{0.48\textwidth}
        \centering
        \includegraphics[width=\linewidth]{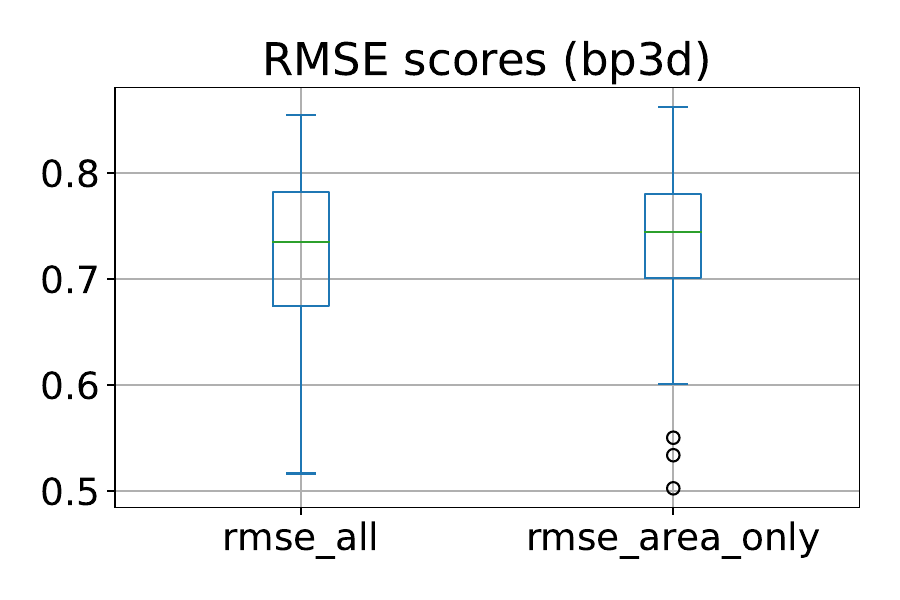}
        \caption{RMSE scores}
    \end{subfigure}%
    \hfill
    \begin{subfigure}[b]{0.48\textwidth}
        \centering
        \includegraphics[width=\linewidth]{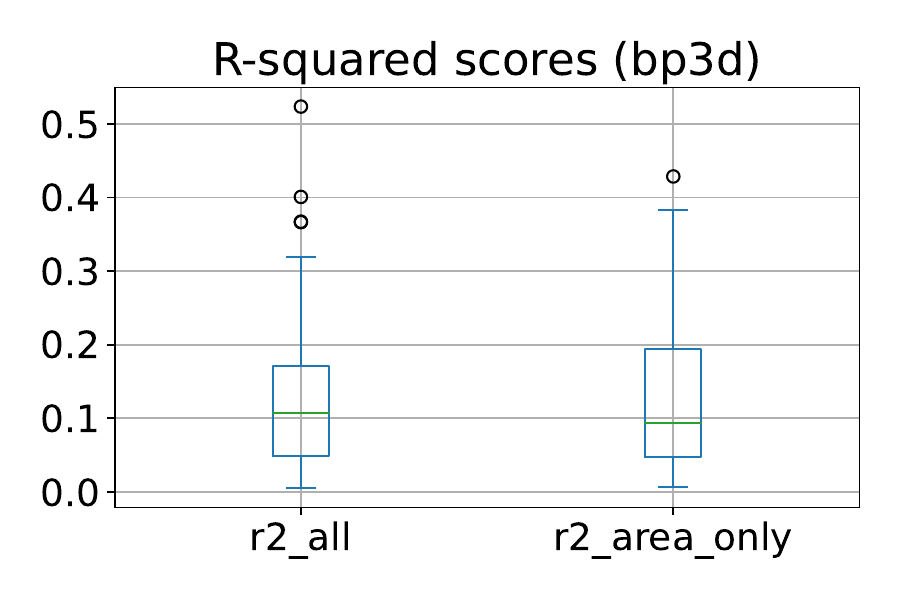}
        \caption{$\text{R}^2$ scores}
    \end{subfigure}


    
    \caption{RMSE and $\text{R}^2$ scores for linear regressions on 25 BP3D samples.}
    \label{fig:lr_bp3d_recommender}
\end{figure}

We trained 100 linear regression models using 25 different data samples for each using all BP3D input features to predict the best hardware configuration. Figure~\ref{fig:lr_bp3d_recommender} shows the performance results of the 100 linear regression models, measured by Root Mean Square Error (RMSE) and $\text{R}^2$ scores. When trained on 25 data samples with all features, the RMSE scores range from $0.5163$ to $0.855$, averaging $0.7256$ with a variation of $0.3387$ RMSE across all 100 models. The $\text{R}^2$ scores indicate that model accuracy varied from $0.48\%$ to $52.36\%$, with an average of $12.83\%$ and a range of $51.88\%$ between the best and worst models. 

\begin{figure}[h!]
    \centering
    \includegraphics[trim={0cm .5cm .4cm 0cm},clip,width=\linewidth]{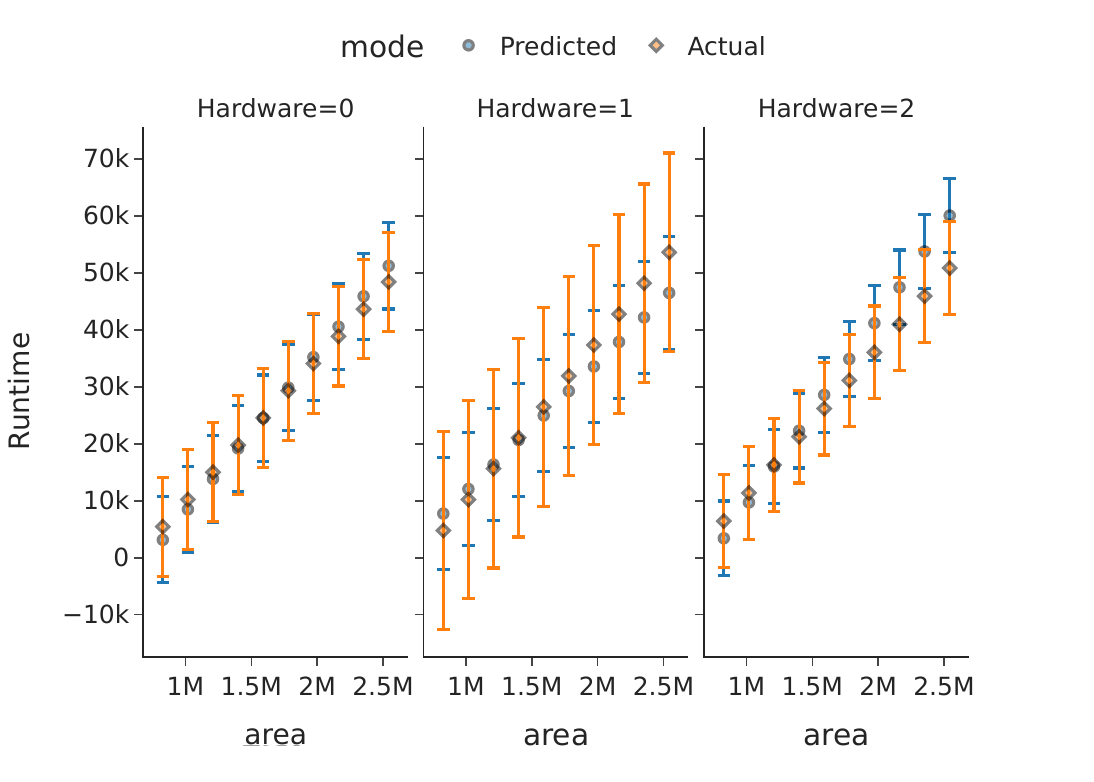}
    \caption{Contextual Bandit vs. Baseline using only \texttt{area} feature. The x-axis presents \texttt{area} in square meters, while the y-axis presents the feature we aim to predict, \texttt{runtime} (in seconds).}
    \label{fig:area_fitting}
\end{figure}

Now, we present the BanditWare results based on Algorithm~\ref{alg:mab}.
We begin by fitting all our data (1316 samples) as the \texttt{baseline} for our prediction. This represents the theoretical best possible model that the contextual bandit can learn. Figure~\ref{fig:area_fitting} shows the fitting for the \texttt{area} feature on each hardware setting $H = \{H0, H1, H2\}$ after running the BanditWare predictor with 100 simulations ($n\_sim = 100$) and 50 rounds ($n\_rounds = 50$). In the plot, we see that our predictor's results closely match the actual values (baseline), although the noise is slightly off.


\begin{figure}[h!]
    \centering
    \begin{subfigure}[b]{0.48\textwidth}
        \centering  
        \includegraphics[trim={.2cm .3cm 2cm 0.4cm},clip,width=\linewidth]{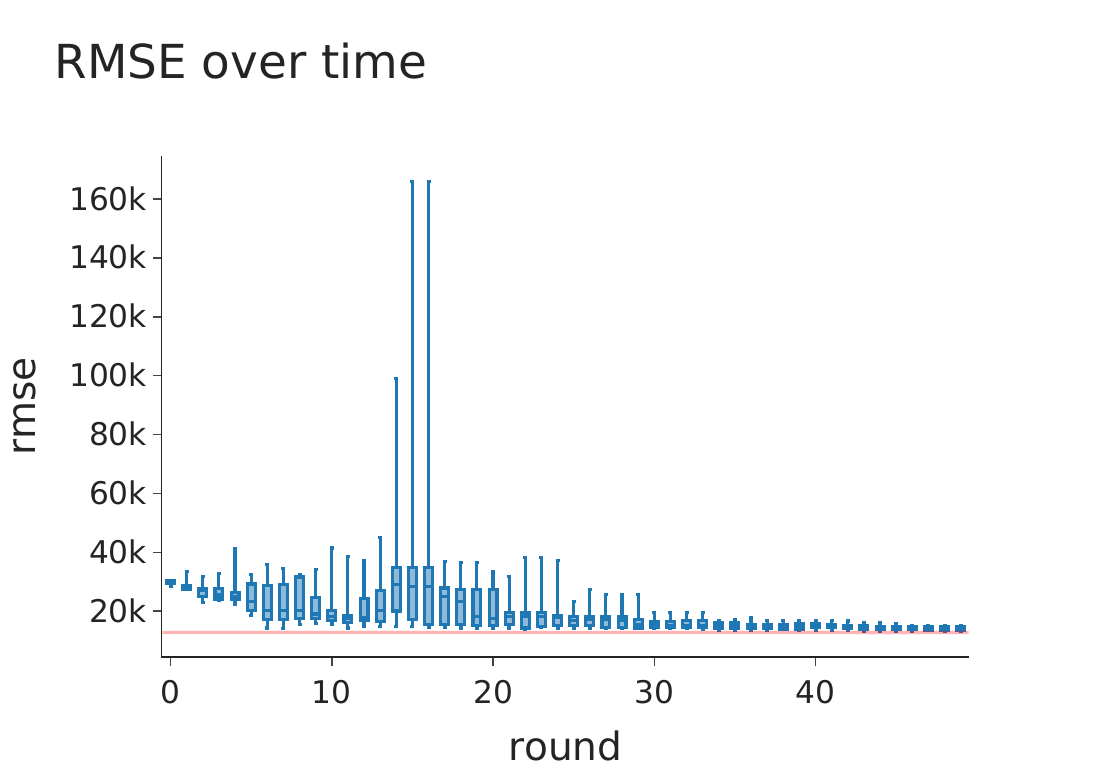}
        \caption{RMSE for BanditWare using \texttt{all} features over time.}
        \label{fig:rmse_all}
    \end{subfigure}%
    \hfill
    \begin{subfigure}[b]{0.48\textwidth}
        \centering
        \includegraphics[trim={.2cm .3cm 2cm 0.4cm},clip,width=\linewidth]{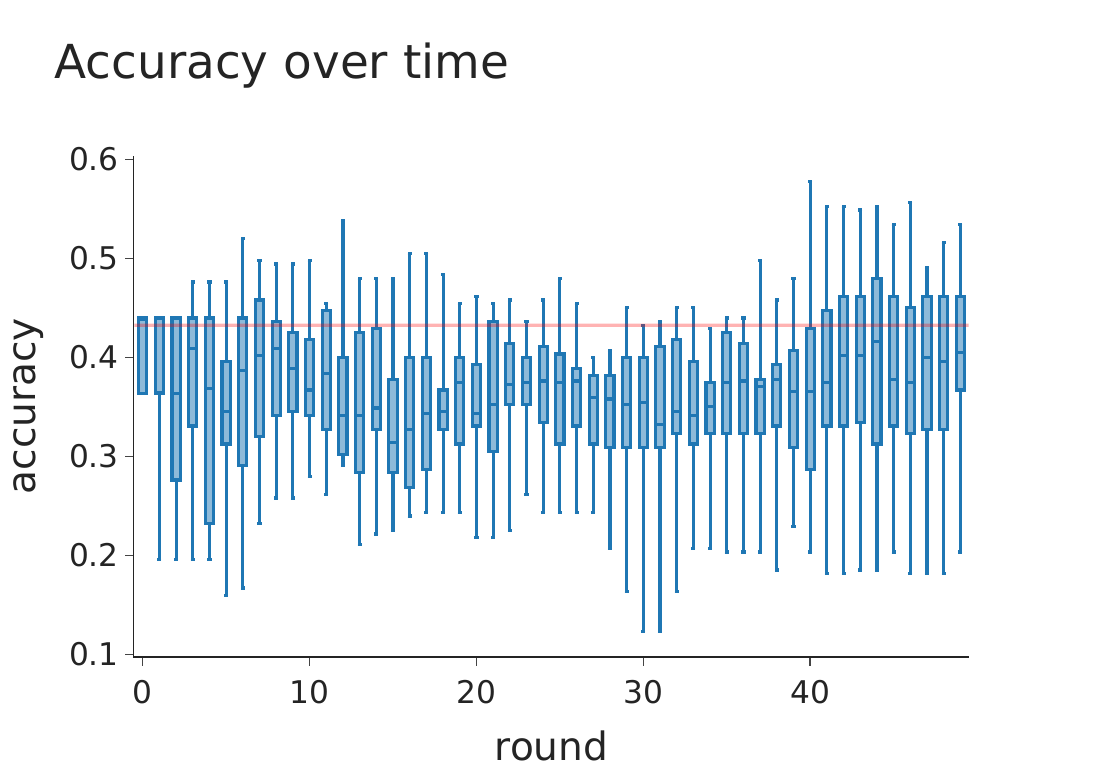}
        \caption{Accuracy of BanditWare using \texttt{all} features over time.}
        \label{fig:accuracy_all}
    \end{subfigure}

    \caption{RMSE and accuracy of BanditWare over time using \texttt{all} features.}
    \label{fig:all}
\end{figure}

Figure~\ref{fig:all} shows that when we use \texttt{all} features, the average RMSE of our framework converges to match the RMSE of all 1316 samples, represented by the orange line, after approximately 25 rounds. The RMSE for the full fit, using all points, is 12257.43. Our predictor achieves an RMSE of $20182.91 \pm 12290.82$ in round 25 and $16493.81 \pm 7078.61$ in round 50, performing 17.90\% and 12.55\% worse than the baseline on average, respectively. We notice that the accuracy of the full fit when using all features has an average of $34.2\%$, which corresponds to the accuracy of randomly choosing the hardware. We believe this occurs because the behaviors of each hardware setting are too similar for our framework to reliably identify the best one in most rounds. In other words, unlike the artificial hardware settings in Experiment 1, there is no clear trade-off between the hardware configurations in this case. Running the application on any of the configurations results in nearly identical runtime. Despite this, the overall results remain promising. Algorithm~\ref{alg:mab} \textbf{is} able to learn a good model for predicting workflow runtime with very few rounds.


\FloatBarrier
\subsection{Experiment 3: Matrix Multiplication}\label{sec-experiments-mm}

To evaluate the effectiveness and limitations of BanditWare on matrix multiplication data, we analyze predictions from two datasets: a complete dataset and a truncated portion. The complete dataset contains 2520 runs with matrices of various sizes ranging from 100 to 12,500. For simplicity, we focus on training using matrix \emph{size} as the predictor, since the other features do not significantly impact the runtime. In this dataset, most runs (1800) involve matrices with \emph{size} $<$ 5000. The secondary dataset is a subset of the complete data, where \emph{size} $\geq$ 5000. This subset likely represents more realistic use cases, where the highest runtime approaches 30 minutes, compared to the maximum of 1 minute for runs with \emph{size} $<$ 5000. Although BanditWare can be used on both datasets, hardware recommendations should emphasize resource allocation rather than speed for runs taking less than a minute, as most hardware configurations perform similarly. Therefore, we also analyze the subset dataset with matrix sizes $\geq$ 5000.

As in the previous use case, we conduct a linear regression experiment to compare the results with BanditWare. We collected results from 100 linear regression recommendation models trained on both the full and truncated matrix multiplication datasets.

\begin{figure}[h!]
    \centering
    \begin{subfigure}[b]{0.48\textwidth}
        \centering
        \includegraphics[width=\linewidth]{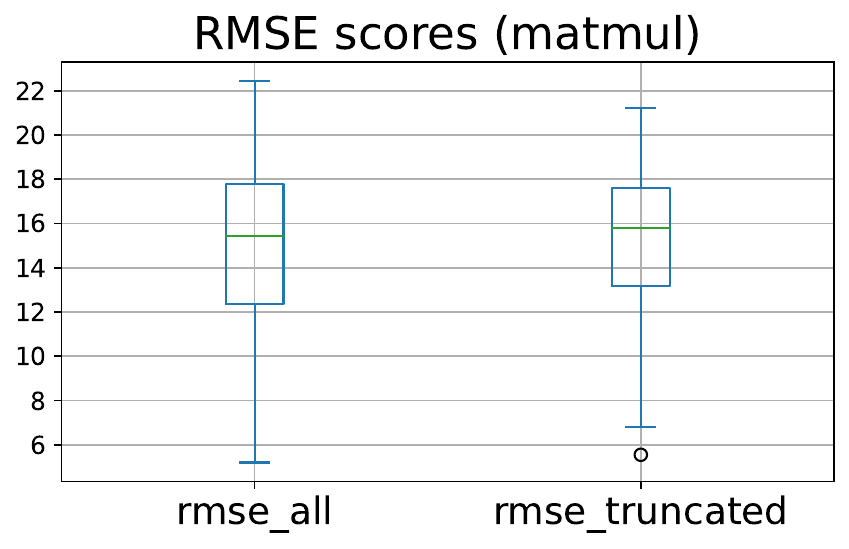}
        \caption{RMSE scores}
    \end{subfigure}
    \hfill
    \begin{subfigure}[b]{0.48\textwidth}
        \centering
        \includegraphics[width=\linewidth]{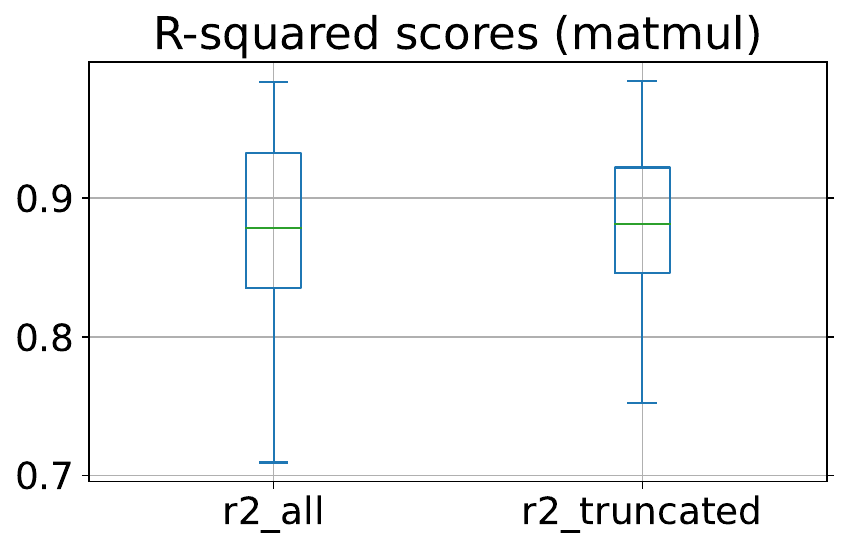}
        \caption{$\text{R}^2$ scores}
    \end{subfigure}


    \caption{Performance metrics for 100 linear regression models trained on matrix squaring data.}
    \label{fig:lr_matmul}
\end{figure}

Figure~\ref{fig:lr_matmul} also shows the distribution of RMSE scores, $\text{R}^2$ scores, and training durations for linear regression recommenders trained on matrix multiplication data. For the full dataset, the lowest RMSE score was $5.1989$, and the highest was $22.4497$, averaging $14.9676$, with a total range of $17.2508$. The $\text{R}^2$ scores ranged from a minimum of $70.9376\%$ to a maximum of $98.3857\%$, averaging $87.6601\%$, with a total range of $27.4481\%$. 

For the truncated dataset with matrix \emph{size} $\geq 5000$, the RMSE scores ranged from $5.5481$ to $21.2297$, averaging $15.0692$ with a range of $15.6816$. The $\text{R}^2$ scores for these models ranged from $75.234\%$ to $97.4758\%$, averaging $88.2434\%$ with a range of $22.2418\%$. The training durations for the truncated dataset models ranged from $1.392$ seconds to $2.423$ seconds, averaging $1.5572$ seconds with a range of $1.031$ seconds.




\begin{figure}[h!]
    \centering
    \begin{subfigure}[b]{0.48\linewidth}
        \centering
        \includegraphics[trim={.3cm 0cm .4cm .3cm},clip,width=\linewidth]{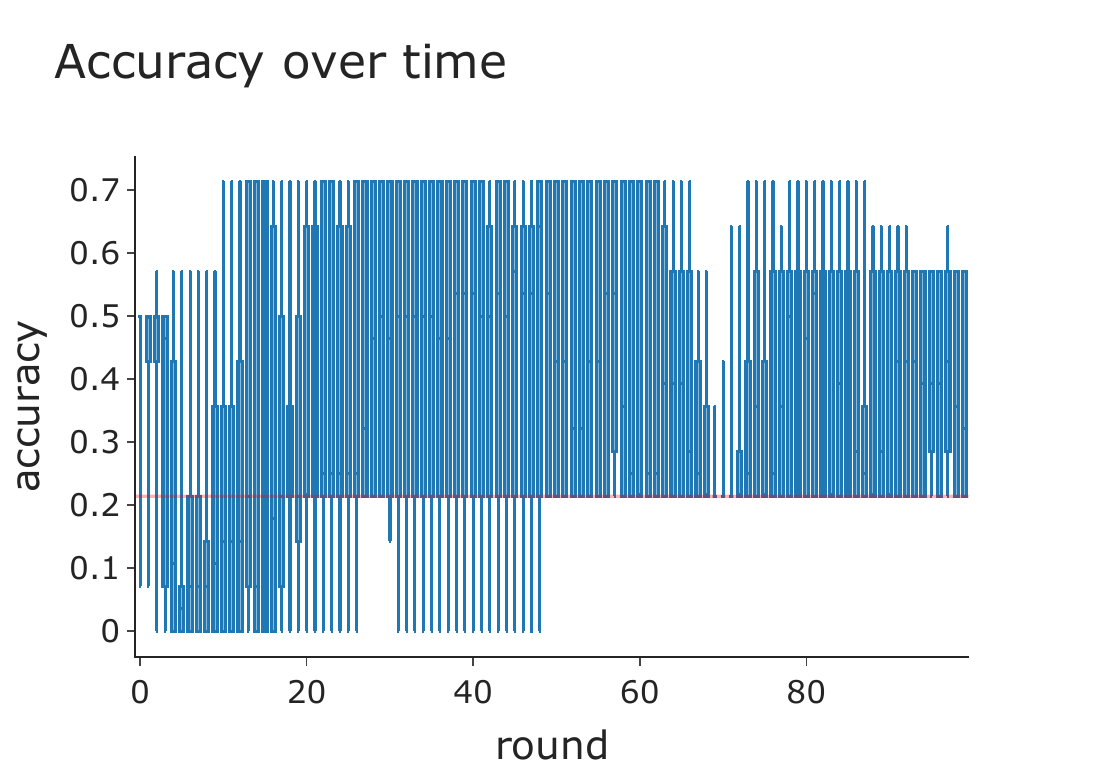}
        \caption{Contextual Bandit accuracy using only \texttt{size} feature with no tolerance for additional slowdown.}
        \label{fig:acc_full_no_tolerance}
    \end{subfigure}
    \hfill
    \begin{subfigure}[b]{0.48\linewidth}
        \centering
        \includegraphics[trim={.3cm .2cm .4cm .7cm},clip,width=\linewidth]{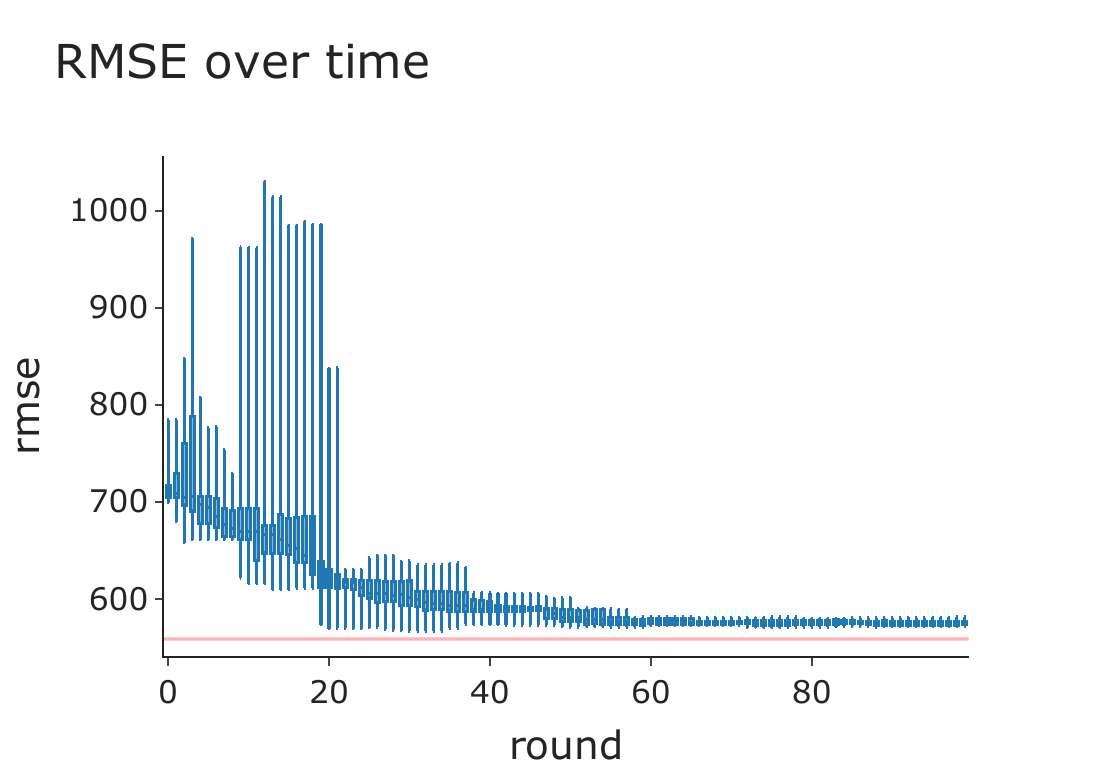}
        \caption{Contextual Bandit RMSE using only \texttt{size} feature with no tolerance for additional slowdown.}
        \label{fig:rmse_full_no_tolerance}
    \end{subfigure}
    \caption{Contextual Bandit accuracy and RMSE on the entire dataset using only \texttt{size} feature for predicting the best hardware with no tolerance for additional slowdown.}
    \label{fig:combined_full_no_tolerance}
\end{figure}



\begin{figure}[h!]
    \centering
    \begin{subfigure}[b]{0.48\linewidth}
        \centering
        \includegraphics[width=\linewidth]{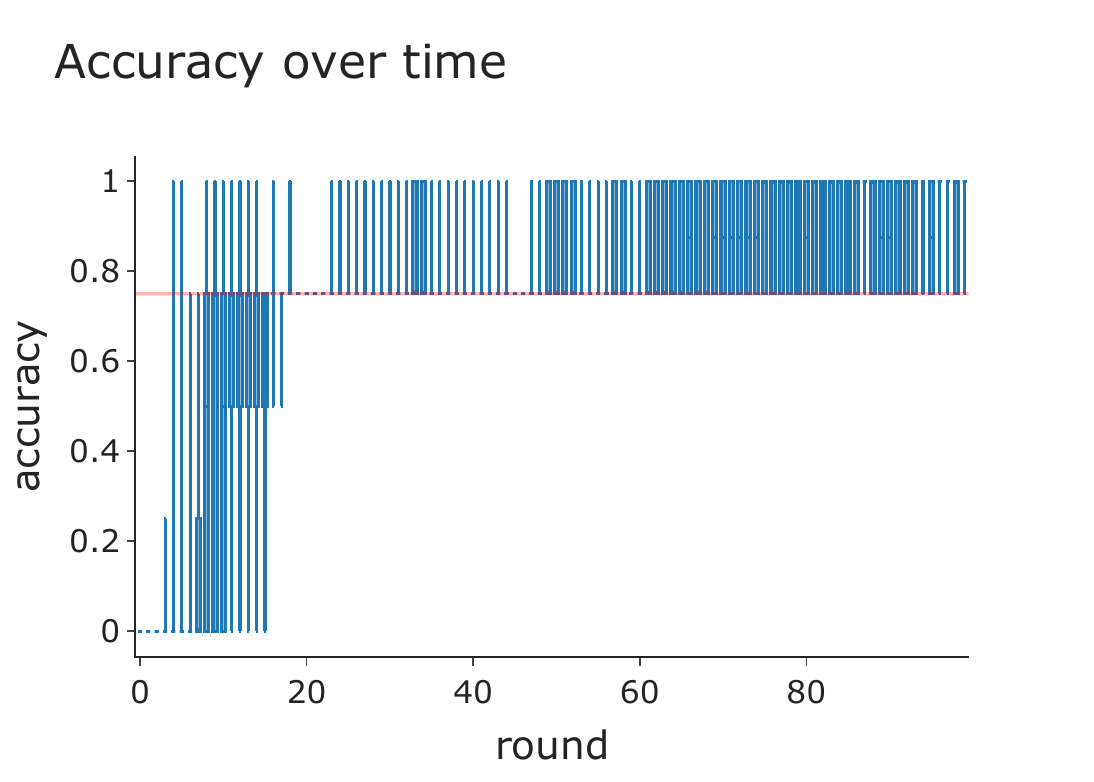}
        \caption{Contextual Bandit accuracy using only \texttt{size} feature with no tolerance for additional slowdown.}
        \label{fig:acc_partial_no_tolerance}
    \end{subfigure}
    \hfill
    \begin{subfigure}[b]{0.48\linewidth}
        \centering
        \includegraphics[width=\linewidth]{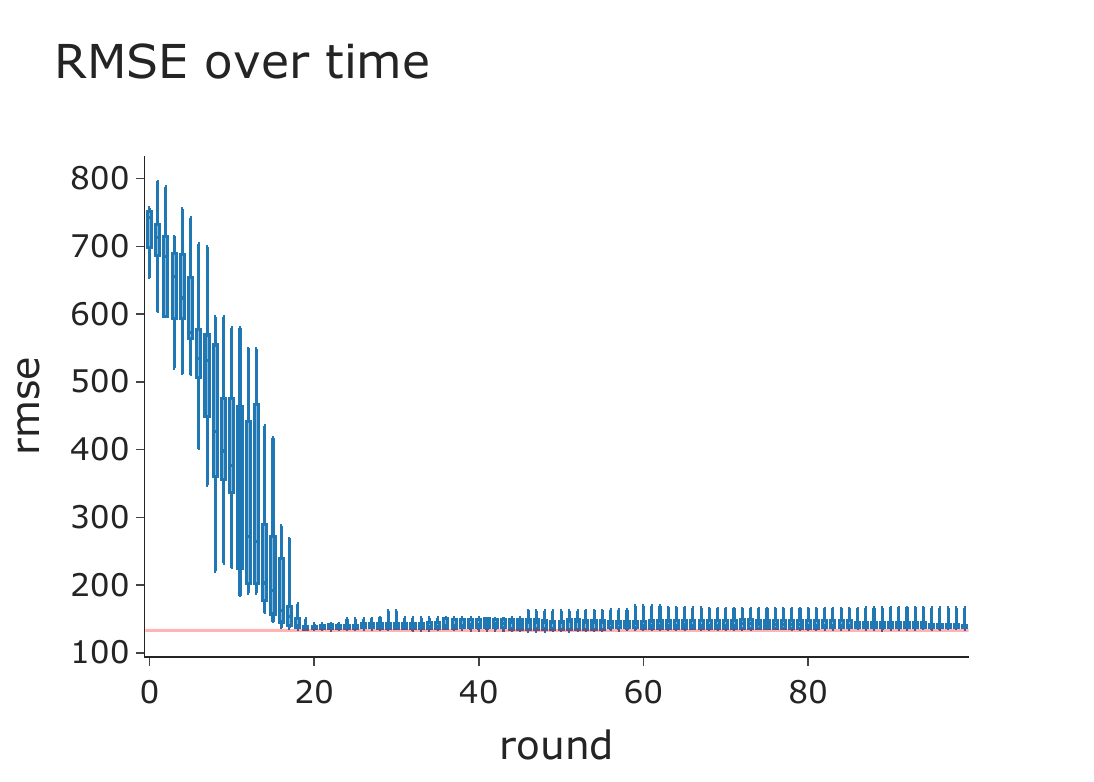}
        \caption{Contextual Bandit RMSE using only \texttt{size} feature with no tolerance for additional slowdown.}
        \label{fig:rmse_partial_no_tolerance}
    \end{subfigure}
    \caption{Contextual Bandit accuracy and RMSE on the subset dataset using only \texttt{size} feature for predicting the best hardware with no tolerance for additional slowdown.}
    \label{fig:combined_partial_no_tolerance}
\end{figure}


\begin{figure}[h!]
    \centering
    \begin{subfigure}[b]{0.48\linewidth}
        \centering
        \includegraphics[width=\linewidth]{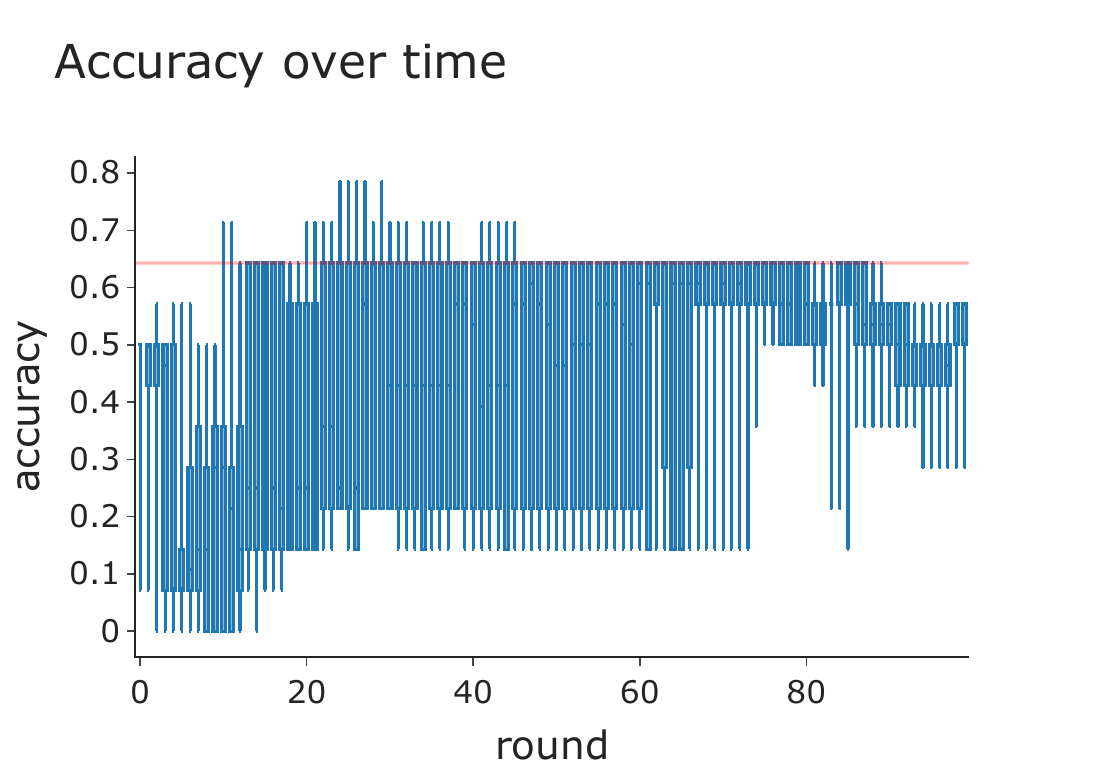}
        \caption{Contextual Bandit accuracy using only \texttt{size} feature with a tolerance of 20 additional seconds.}
        \label{fig:acc_full_w_tolerance}
    \end{subfigure}
    \hfill
    \begin{subfigure}[b]{0.48\linewidth}
        \centering
        \includegraphics[width=\linewidth]{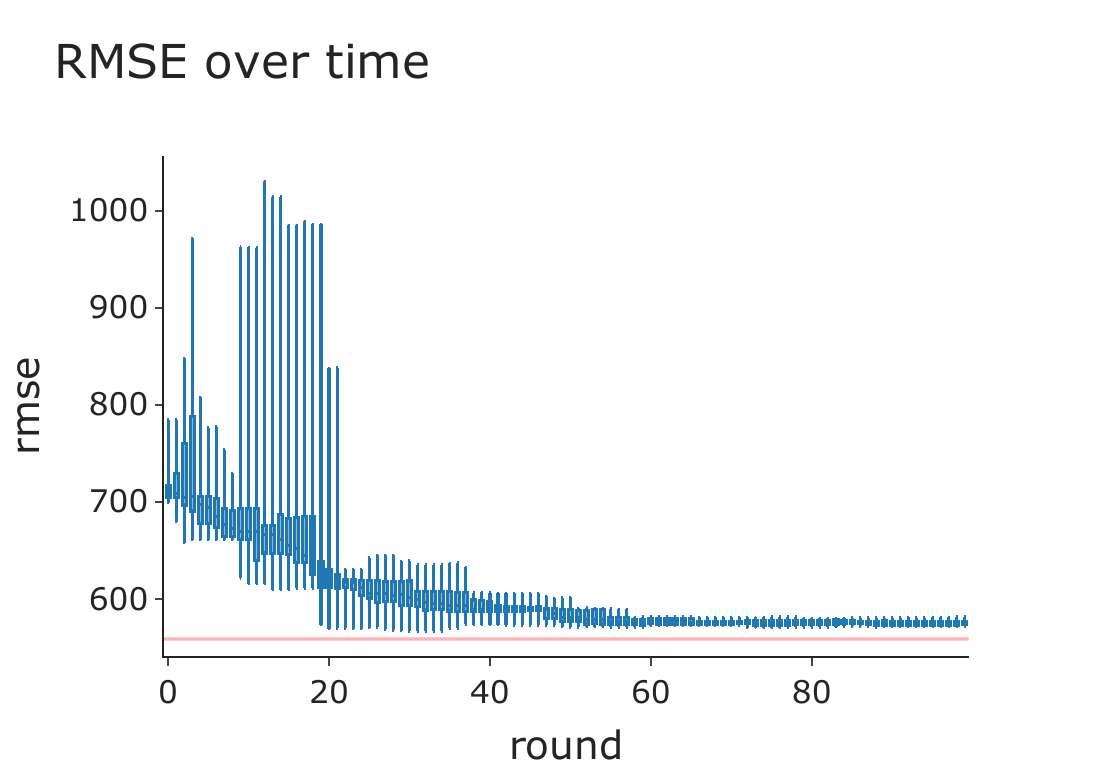}
        \caption{Contextual Bandit RMSE using only \texttt{size} feature with a tolerance of 20 additional seconds.}
        \label{fig:rmse_full_w_tolerance}
    \end{subfigure}
    \caption{Contextual Bandit accuracy and RMSE on the entire dataset using only \texttt{size} feature for predicting the best hardware with a tolerance of 20 additional seconds.}
    \label{fig:combined_full_w_tolerance}
\end{figure}


\begin{figure}[h!]
    \centering
    \begin{subfigure}[b]{0.48\linewidth}
        \centering
        \includegraphics[width=\linewidth]{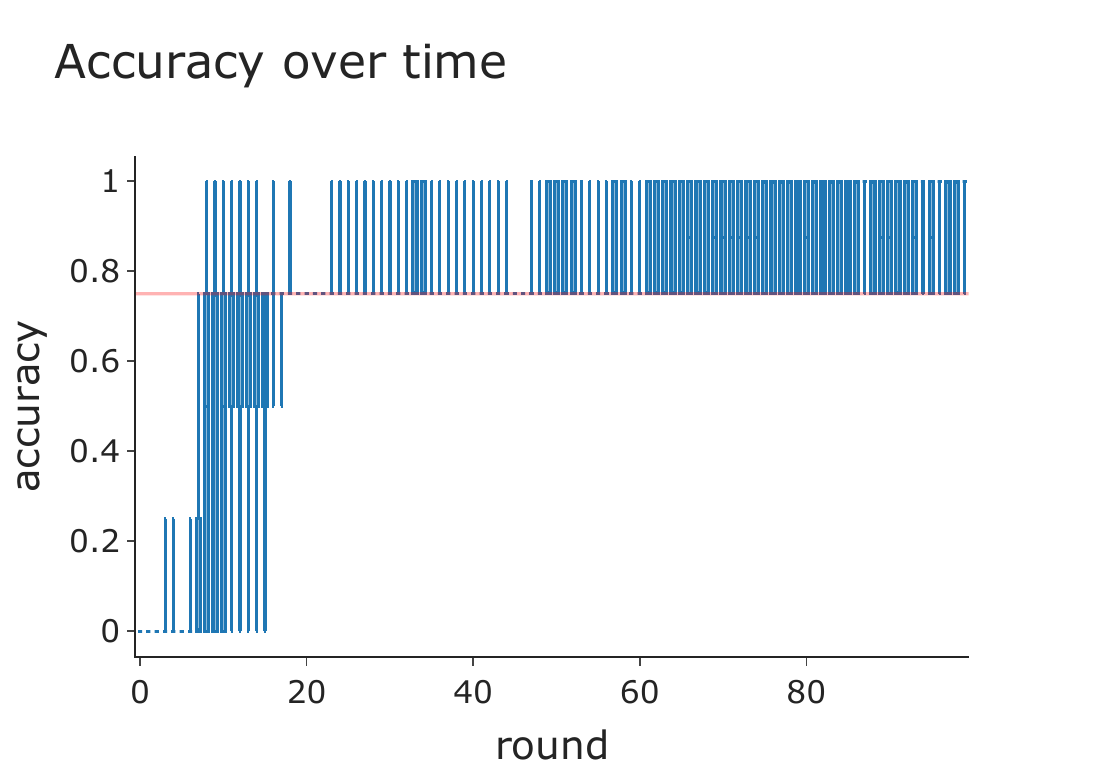}
        \caption{Contextual Bandit accuracy using only \texttt{size} feature with a tolerance of 5\% slowdown.}
        \label{fig:acc_partial_w_tolerance}
    \end{subfigure}
    \hfill
    \begin{subfigure}[b]{0.48\linewidth}
        \centering
        \includegraphics[width=\linewidth]{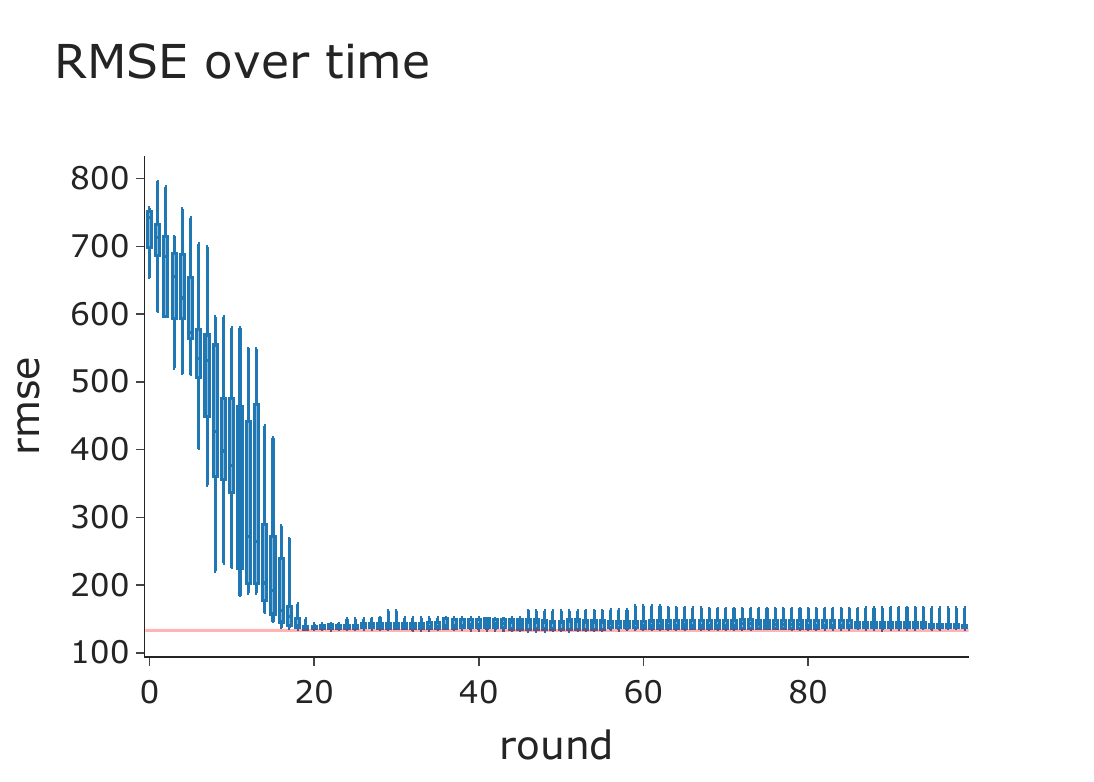}
        \caption{Contextual Bandit RMSE using only \texttt{size} feature with a tolerance of 5\% slowdown.}
        \label{fig:rmse_partial_w_tolerance}
    \end{subfigure}
    \caption{Contextual Bandit accuracy and RMSE on the subset dataset using only \texttt{size} feature for predicting the best hardware with a tolerance of 5\% slowdown.}
    \label{fig:combined_partial_w_tolerance}
\end{figure}

We ran Algorithm~\ref{alg:mab} on both datasets (full dataset and dataset with \emph{size} $\geq$ 5000) and observed that, as shown in Figure~\ref{fig:acc_full_no_tolerance}, the algorithm struggles to predict the best hardware on the full dataset, where runtimes are relatively small. Here, BanditWare has a full fit accuracy of approximately 0.3, compared to a random guess accuracy of 0.2 among the five hardware options. However, when we train on the subset dataset with larger matrix sizes (\emph{size} $\geq$ 5000), the accuracy increases significantly, reaching nearly 0.8, as shown in Figure~\ref{fig:acc_partial_no_tolerance}. This behavior is not necessarily problematic, since perfect recommendations based on minimum runtime are less critical for short runtimes. In such cases, optimizing for resource efficiency is more important than minimizing runtime. Therefore, specifying the tolerance\_seconds parameter for BanditWare is beneficial, allowing for a slight increase in runtime in exchange for lower resource consumption.

When we set the tolerance to 20 additional seconds, as shown in Figure~\ref{fig:acc_full_w_tolerance}, we observe a significant improvement in accuracy while selecting less resource-intensive hardware. For the subset dataset, where most runs take several minutes, the tolerance\_seconds parameter is less impactful, and using tolerance\_ratio becomes more relevant to allow for larger runtime differences. BanditWare offers the flexibility to specify both tolerance\_ratio and tolerance\_seconds, ensuring optimal performance across different runtime scenarios. Adding a 5\% tolerance for slowdown in the subset dataset, as shown in Figure~\ref{fig:acc_partial_w_tolerance}, allows for high accuracy while selecting more efficient hardware configurations. If minimizing runtime is crucial, setting the tolerance parameters to zero will prevent hardware configurations that increase runtime. Overall, BanditWare performs significantly better on the subset dataset, where most runtimes exceed one minute. By tolerating some runtime slowdown, BanditWare achieves higher accuracy and selects more resource-efficient hardware configurations, addressing the performance issues observed in datasets with large size variations.

\section{Conclusion} \label{sec-conclusion}

We presented BanditWare, a lightweight recommendation framework based on a contextual bandit that suggests the best-fitting hardware to run an application. Our results showed that, in just 25 rounds, our approach learns a model that performs only 17.90\% worse than the theoretically best possible. In some experiments the framework's accuracy was impacted by the homogeneity of the hardware available. We also demonstrated that BanditWare's tolerance functionality mitigates trade-offs between performance consistency and speed when compared to a simple linear regression recommendation modeling system using the same data. Furthermore, as illustrated Figure~\ref{fig:motivation} (from Experiment 1), our framework performs well when there is greater variation in the hardware settings. 

For future work, we plan to explore different and more complex contextual bandit algorithms, along with a broader variety of hardware configurations. We will also experiment with additional applications, including large language models (LLMs), enabling us to incorporate GPU information into hardware recommendations. We aim to adapt BanditWare to support multiple parameter minimization, and we plan on monitoring more performance metrics, such as communication latency and scheduling overhead, for a more well-rounded tool that can be used at scale.

\section*{Acknowledgement} This work was supported by NSF award 2134904 for BurnPro3D and NSF award 2333609 for the National Data Platform and Schmidt Sciences. The Nautilus Kubernetes Cluster of the National Research Platform, partially funded by NSF awards 2112167 and 2120019, was also used for this research. The authors thank the WIFIRE and WorDS teams at the San Diego Supercomputer Center for their collaboration.



\bibliographystyle{IEEEtran}
\bibliography{references}


\end{document}